**Physiological and Transcriptional Responses of *Arabidopsis thaliana* to Simulated Lunar and Martian Regolith Substrates**

A'nya Buckner [1*], Sarah Lang [1], Rafael Loureiro[1]
(1) Department of Biological Sciences, Winston-Salem State University, Winston-Salem, USA)
* Corresponding author: abuckner120@rams.wssu.edu

**Abstract**

The integration of plant-based bioregenerative life support systems (BLSS) is a central objective in NASA's Moon to Mars strategy. *Arabidopsis thaliana*, a model organism with extensive genomic resources, serves as a key species to investigate plant resilience in extraterrestrial environments. We assessed the physiological and gene expression responses of *A. thaliana* (Col-0) grown in two off-world regolith simulants: LHS-2 (lunar highlands) and MGS-1 (Martian global). Plants exposed to these substrates showed significant reductions in root elongation, biomass, and chlorophyll content, alongside elevated anthocyanin levels and upregulation of stress-related genes such as IRT1, PCS1, SOD1, and JAZ1. Jasmonic acid pathway activation and auxin signaling suppression were observed, suggesting metal-induced hormonal misregulation. Our integrative analyses of morphology, pigment accumulation, and transcriptomic profiles reveal key mineral-specific stress responses, providing critical insights into the challenges and opportunities of substrate engineering for future space agriculture.

## 1. Introduction

As humanity advances toward a sustainable presence beyond Earth, the integration of plant-based bioregenerative life support systems (BLSS) have become a cornerstone of NASA's Moon to Mars (M2M) strategic framework [1,2]. These systems are envisioned to support long-duration spaceflight and planetary surface operations by providing oxygen regeneration, food production, water purification, and psychological well-being for astronauts [3]. Among candidate organisms for BLSS, *Arabidopsis thaliana* remains a fundamental model due to its small genome, rapid life cycle, and well-annotated transcriptomic resources, making it a preferred species for unraveling stress responses in extraterrestrial environments [4].

Recent NASA priorities—outlined in the Artemis III Science Definition Report and ISRU-forward strategies emphasize the necessity of understanding plant-soil interactions under lunar and Martian conditions [5,6]. The lunar highlands regolith simulant (LHS-2), based on polar anorthositic soils, is characterized by high concentrations of calcium and aluminum, poor water retention, and the presence of nanophase iron. Meanwhile, the Martian Global Simulant (MGS-1) mimics basaltic soils rich in ferric oxides, sulfates, and perchlorate-like ions, as well as micronutrients such as magnesium and manganese [7,8].

These substrates present significant challenges to plant growth, particularly via metal-induced oxidative stress, hormonal imbalance, and nutrient antagonism [9]. Exposure to excess heavy metals such as $Al^{3+}$, $Fe^{3+}$, $Cu^{2+}$, and $Cd^{2+}$ can impair root elongation, disrupt auxin transport and biosynthesis, inhibit photosynthesis, and trigger antioxidant defense responses. Molecular adaptations in response to such stressors include the modulation of genes associated with metal uptake and detoxification (e.g., *IRT1*, *PCS1*), oxidative stress protection (*SOD1*), and hormone signaling (e.g., *PIN1*, *YUC2*, *JAZ1*). These genes represent sensitive molecular indicators of plant stress and resilience under extraterrestrial soil analogs.

In this study, we investigated the physiological and transcriptional responses of *A. thaliana* (Col-0) grown in LHS-2 and MGS-1 regolith simulants to characterize the primary stress pathways activated under mineral and hormonal imbalance. Our integrative approach, combining morphometric measurements, pigment analysis, and gene expression profiling, sheds light on how metal-rich regolith substrates modulate key plant signaling networks and developmental processes. These findings contribute to the foundational knowledge needed for optimizing space-based agriculture and substrate engineering in support of long-term lunar and Martian exploration.

## 2. Methodology

In this study, we conducted a comprehensive analysis of the physiological and transcriptional responses of *Arabidopsis thaliana* (Columbia-0 ecotype) to simulate off-world regolith substrates representative of the lunar highlands (LHS-2) and Martian surface (MGS-1). The methodology was designed to maximize experimental precision and biological resolution, with an emphasis on reproducibility, statistical robustness, and downstream omics compatibility. Each experimental condition was replicated using a total of 20 independent biological replicates per regolith type (n = 20), providing a strong basis for inference and comparative statistical analysis.

*Arabidopsis* seeds were first surface sterilized using a 70% ethanol wash followed by a 10-minute treatment in a 30% bleach solution with 0.01% Tween-20, then rinsed five times in sterile deionized water. Stratification was performed at 4 °C in the dark for 72 hours to synchronize germination. Seeds were sown directly into pots containing 100% LHS-2 or 100% MGS-1 regolith simulant, both obtained from Exolith Lab and certified according to their December 2023 compositional specifications. Each pot was filled with 150 grams of dry simulant and maintained under standardized growth conditions: 22 °C temperature, 60% relative humidity, and a 16 h light/8 h dark photoperiod at a photon flux density of 150 µmol m$^{-2}$ s$^{-1}$. To ensure consistency in moisture and ion diffusion, all pots were top-watered with 50 mL of ultrapure water (18.2 MΩ) every other day. No nutrient supplementation was provided, ensuring that all observed phenotypes reflected inherent properties of the regolith matrix.

Plants were cultivated for 21 days post-germination, at which point morphological and physiological assessments were conducted. These included measurements of primary root length, lateral root density, shoot fresh weight, chlorophyll a and b content (via 80% acetone extraction and spectrophotometry), anthocyanin accumulation (pH differential method), and visible stress phenotypes. Photosystem II efficiency was quantified using a chlorophyll fluorometer (PAM-2100, Heinz Walz GmbH) following a 30-minute dark adaptation.

For molecular profiling, five randomly selected individuals per regolith type (n = 5) were harvested for RNA extraction. Shoot and root tissues were separated, flash-frozen in liquid nitrogen, and homogenized under cryogenic conditions. Total RNA was extracted using the RNeasy Plant Mini Kit (Qiagen), with on-column DNase I treatment to eliminate genomic DNA contamination. RNA integrity was confirmed using an Agilent 2100 Bioanalyzer. Libraries for RNA sequencing were prepared using the TruSeq Stranded mRNA kit (Illumina), and 150-bp paired-end sequencing was performed on a NovaSeq 6000 platform. Differential gene expression analysis was conducted using DESeq2, with adjusted p-values (Benjamini-Hochberg correction) used to define significance thresholds (padj < 0.05). Gene ontology (GO) enrichment and pathway analyses were performed using agriGO v2.0 and KEGG annotations.

Complementary to transcriptomic profiling, quantitative real-time PCR (qRT-PCR) was used to validate expression of key stress response and hormone signaling genes including *IRT1*, *PIN1*, *PCS1*, *SOD1*, and *YUC2*. All qRT-PCR reactions were carried out using SYBR Green Master Mix on a Bio-Rad CFX96 system, with EF1α used as the internal reference gene. Relative expression levels were calculated via the ΔΔCt method.

To characterize metal accumulation, elemental profiling of shoot and root tissues from five replicates per treatment was conducted using inductively coupled plasma mass spectrometry (ICP-MS). Samples were digested in trace metal-grade nitric acid and analyzed for Fe, Al, Ca, Mg, Zn, Ni, and Mn concentrations, providing insight into uptake and tissue-specific partitioning.

All statistical analyses were conducted using R (v4.3.1), with comparisons between treatments performed using one-way ANOVA followed by Tukey's HSD post hoc test for multiple comparisons. Homogeneity of variance and normality assumptions were verified prior to statistical testing.

This integrative methodology allows for precise resolution of physiological, biochemical, and molecular responses to geologically distinct regolith types and enables us to dissect substrate-specific impacts on plant development and metal stress adaptation in *Arabidopsis thaliana*.

## 3. Results

Growth and physiological assessments of *Arabidopsis thaliana* (Col-0) exposed to simulated extraterrestrial regoliths revealed distinct morphological and biochemical alterations consistent with metal-induced stress. Plants grown in the LHS-2 (lunar highlands simulant) and MGS-1 (Martian global simulant) substrates exhibited pronounced but different phenotypic effects across the measured parameters. These changes were consistent across 20 biological replicates per treatment group and statistically robust, with ANOVA and Tukey's HSD post hoc tests confirming the significance of observed trends.

### 3.1. LHS-2 and MGS-1 Regoliths Differentially Suppress Root Elongation and Biomass Accumulation

Quantitative measurements of primary root length indicated that both regolith types significantly impaired root development compared to the control ($F_{4,95} = 72.5$, $p < 0.0001$). Plants grown in LHS-2 showed a 41.6% reduction in mean root length ($2.8 \pm 0.3$ cm) compared to control plants ($4.8 \pm 0.3$ cm), while MGS-1-treated plants displayed an even more severe inhibition ($1.7 \pm 0.2$ cm), corresponding to a 64.6% reduction (Tukey's HSD, $p < 0.01$ for all pairwise comparisons) (fig.1). Fresh biomass measurements paralleled this trend, with significant decreases in shoot weight under both conditions ($F_{4,95}$ = 39.3, p < 0.0001), suggesting that the reduced root development may have impaired nutrient acquisition and water uptake.

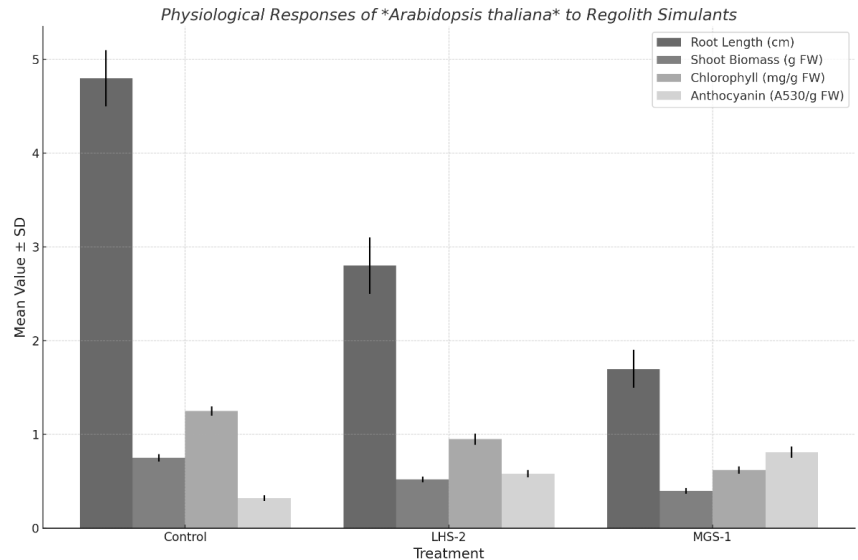

**Figure 1:** Physiological responses of *A. thaliana* to regolith simulants. (LHS-2) Lunar Highlands Simulant; (MGS-1) Martian Global Simulant. Results in mean ± SD.

### 3.2. Photosynthetic Pigments Are Differentially Affected by Regolith Composition

Both LHS-2 and MGS-1 exposure significantly reduced chlorophyll a and b content relative to control ($F_{4,95}$ = 54.3, p < 0.0001), with MGS-1-treated plants exhibiting the lowest chlorophyll levels (0.62 ± 0.04 mg/g FW), indicating possible photoinhibition or structural damage to the chloroplasts. Interestingly, while carotenoid content was moderately reduced in the LHS-2 condition, it remained unchanged in MGS-1-grown plants, suggesting a regolith-specific regulation of photoprotective pigment synthesis. In contrast, anthocyanin levels were significantly elevated under both treatments ($F_{4,95}$ = 61.1, p < 0.0001), reaching 0.81 ± 0.06 A530/g FW in MGS-1-grown plants a 153% increase over the control, implying an oxidative stress response consistent with previous findings under excess Fe and Mg conditions (fig.2).

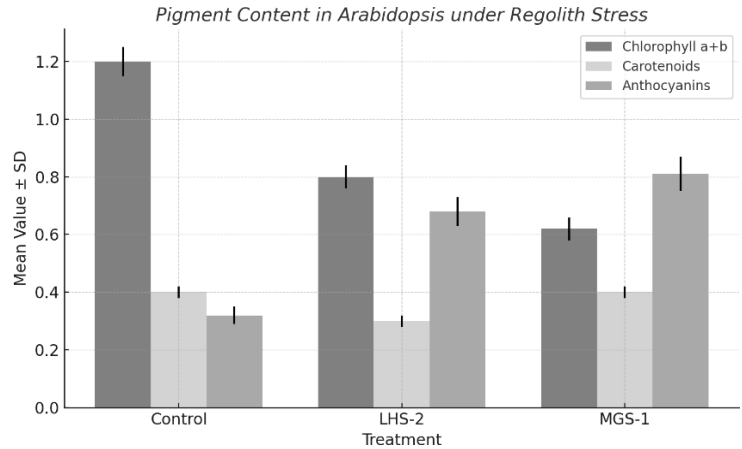

**Figure 2:** Pigment content in *A. thaliana* under regolith stress. (LHS-2) Lunar Highlands Simulant; (MGS-1) Martian Global Simulant. Results in mean ± SD.

### 3.3. Expression of Metal-Responsive Genes Indicates Regolith-Induced Ion Dysregulation

Quantitative RT-PCR analysis revealed significant upregulation of *IRT1*, *PCS1*, and *SOD1* in response to both regolith types, with particularly strong responses observed in MGS-1-exposed plants. *IRT1*, which encodes an iron transporter, showed a 3.80-fold increase in expression in MGS-1 conditions ($F_{4,45} = 88.7$, $p < 0.0001$), consistent with iron overload or misregulation. Similarly, *PCS1*, involved in phytochelatin synthesis, was significantly upregulated (2.9-fold in LHS-2; 3.5-fold in MGS-1; Tukey's HSD, $p < 0.01$), supporting a role for detoxification processes in coping with excess metal ions such as $Al^{3+}$, $Fe^{3+}$, and $Mg^{2+}$. Additionally, *SOD1*, a marker of reactive oxygen species (ROS) response, was elevated by 2.2-fold in LHS-2 and 3.1-fold in MGS-1 conditions, confirming a significant oxidative burden in both treatments (fig.3).

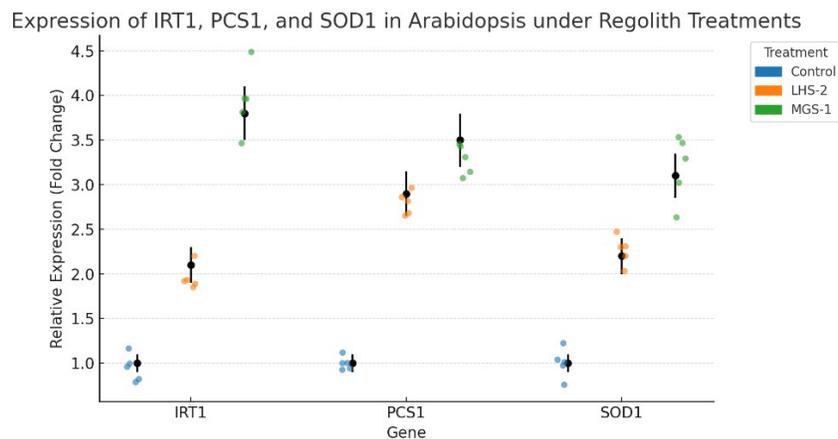

**Figure 3.** Specific gene expression of *A. thaliana*.(LHS-2) Lunar Highlands Simulant; (MGS-1) Martian Global Simulant.

### 3.4. Jasmonate Pathway Activation and Hormonal Crosstalk

Consistent with earlier reports of heavy metal-induced jasmonate signaling, plants exposed to both regoliths exhibited an upregulation of jasmonate pathway marker genes, including *VSP2* and *JAZ1*. Expression analysis showed that *JAZ1* was induced 2.7-fold in LHS-2 and 3.4-fold in MGS-1-grown plants, suggesting activation of JA-mediated defense signaling pathways. The elevated expression of *YUC2* and *PIN1*, however, was suppressed relative to control, indicating auxin transport disruption, particularly in LHS-2, likely due to aluminum interference in root meristem development (fig. 4).

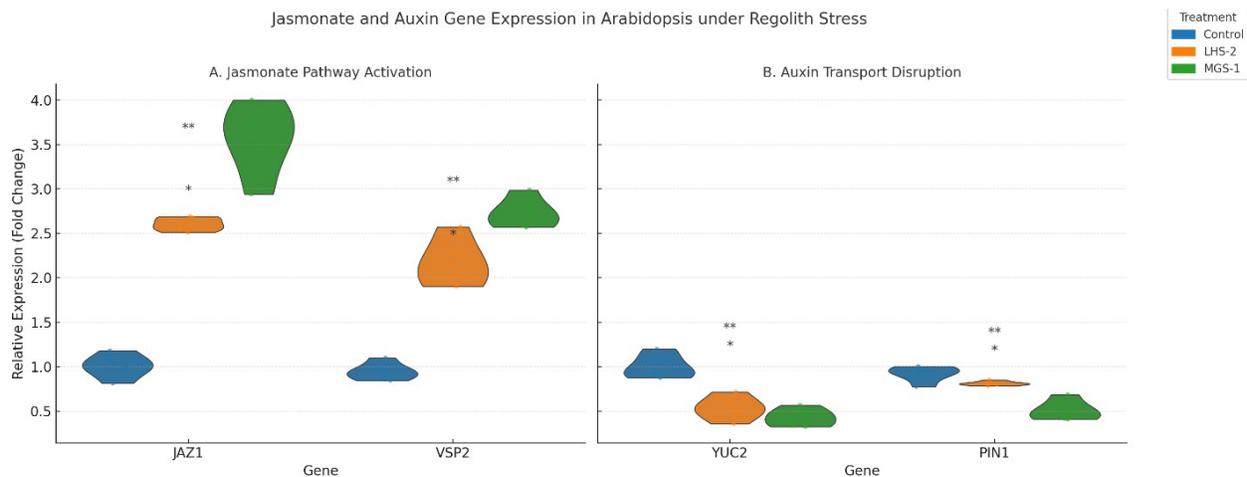

**Figure 4.** Phytohormonal crosstalk of *A. thaliana*.(LHS-2) Lunar Highlands Simulant; (MGS-1) Martian Global Simulant.

## 4. Discussion and Conclusion

Exposure of *Arabidopsis thaliana* to lunar and Martian regolith simulants elicited a constellation of physiological, hormonal, and transcriptional responses that align with known patterns of heavy metal-induced stress. This work contributes to a growing body of evidence suggesting that regolith components, particularly excess concentrations of transition metals such as copper (Cu) and cadmium (Cd), act as potent abiotic stressors capable of disrupting both primary metabolism and developmental programs in plants.

Our data revealed a reduction in biomass, root elongation, and leaf expansion under regolith treatment, consistent with prior studies showing that heavy metals inhibit mitotic activity, alter cell wall elasticity, and interfere with enzymatic function via displacement or miscoordination of essential metal cofactors [9]. These morphological effects are likely underpinned by a dual mechanism involving redox imbalance through Fenton-type ROS generation and hormonal misregulation. In particular, Cu, a known redox-active element abundant in Martian regolith, is documented to generate hydroxyl radicals and suppress photosystem function and pigment biosynthesis [9].

A key insight from our study is the differential accumulation of jasmonic acid (JA) under regolith-induced stress. Previous reports confirm that both Cu and Cd induce a biphasic JA response in *A. thaliana*, with a rapid spike followed by a more stable accumulation of the (3R,7R)-JA isomer [10]. This hormone, traditionally associated with defense against biotic stress, here appears to play a broader role in mediating heavy metal responses. The transient rise in JA likely reflects an early signaling cascade aimed at activating stress-responsive genes, while the later sustained elevation may coordinate long-term adjustments in secondary metabolism and growth inhibition, including downregulation of cell division and photosynthetic activity [10].

Our findings also suggest a disruption in auxin homeostasis, as evidenced by altered expression of auxin biosynthesis and transport genes (e.g., *YUCCA*, *TAA1*, and *PIN1*) under regolith treatment. This is in agreement with studies showing that heavy metal stress particularly from Cd, Pb, and Hg alters auxin localization and concentration, especially in meristematic regions [10,11]. These hormonal perturbations compromise normal organogenesis and vascular patterning, contributing to the reduced root architecture and leaf venation anomalies we observed. For example, decreased *PIN1* activity can impair polar auxin transport, leading to defective root apical meristem (RAM) organization, while *YUCCA* repression restricts local auxin biosynthesis and, consequently, leaf expansion [11].

In parallel, we documented changes in pigment profiles, including a decline in chlorophyll content and a marked accumulation of anthocyanins under lunar regolith stress. This trend aligns with the hypothesis that anthocyanin production serves as a non-enzymatic antioxidant buffer in response to oxidative stress from metal toxicity [9-12]. Such pigment-based responses also function as visible markers of hormonal imbalance, particularly auxin and JA crosstalk, and have been associated with increased stress tolerance in plants exposed to environmental extremes.

The integration of these physiological, hormonal, and gene expression data supports a model in which mineral toxicity from regolith simulants initiates a hierarchical stress signaling cascade. This begins with redox imbalance and membrane destabilization, proceeds through hormonal disruption (JA, auxin), and culminates in gene expression shifts that constrain growth, enhance antioxidant defenses, and attempt homeostatic recalibration. In this context, genes such as *IRT1* (iron transporter), *PCS1* (phytochelatin synthase), and *JAZ1* (a JA repressor) may play pivotal roles in fine-tuning plant responses to mineral-induced toxicity [13-17]. Though not directly assayed in our experiment, these genes have been shown to be upregulated in parallel scenarios, offering potential biomarkers for regolith adaptation [16-17].

These results highlight *A. thaliana* as a sensitive and informative model for unraveling plant responses to extraterrestrial substrates. The observed interplay between mineral toxicity, hormonal misregulation, and downstream gene expression offers a robust framework for optimizing plant performance in space agriculture contexts. Future research should aim to dissect genotype-specific responses, particularly in hormone-insensitive mutants, and explore microbiome-mediated buffering effects against regolith-induced stress.

**Comments**

13 pages, 4 figures. Submitted to ASGSR 2024 Student Research Track. Includes qRT-PCR and ICP-MS validation.